# Dielectric Sensing in ε-Near-Zero Narrow Waveguide Channels

Andrea Alù, and Nader Engheta[*]

Department of Electrical and Systems Engineering, University of Pennsylvania

Philadelphia, PA 19104, U.S.A.

*We exploit here the dramatic field enhancement caused by energy squeezing and tunneling (i.e., "supercoupling") in metamaterial-inspired ultranarrow waveguide channels with near zero effective permittivity in order to sense small permittivity variations in a tiny object. The supercoupling effect is accurately modeled analytically and closed form expressions are derived to describe the presence of defects or permittivity perturbations along the channel. Applications for tailoring its pass-band frequency and for high-Q sensing are proposed at microwave frequencies.*

The possibility of squeezing energy through sub-wavelength apertures and channels has been recently investigated in several systems and setups involving plasmonic materials and metamaterials (see, e.g., [1]-[3]). Somewhat analogous to these findings, the idea of metamaterial supercoupling, i.e., energy squeezing and tunneling through sub-wavelength narrow waveguide channels of arbitrary length and narrow height filled with epsilon-near-zero (ENZ) metamaterials, has been put forward in a recent letter [4]. Further theoretical results and experimental realization of this anomalous phenomenon have been obtained [5]-[7], demonstrating how it is possible to induce such tunneling in various waveguide configurations at microwave frequencies. In particular, we have reported in [5]-[6] how the same tunneling phenomenon may be achieved inside a simple hollow rectangular channel of narrow height that operates near the cutoff frequency of its dominant $TE_{10}$ mode, thus making use of its natural dispersion to mimic the response of ENZ materials [8]-[9]. This may lead to resonant transmission and vanishingly small phase delay through arbitrarily narrow waveguide channels, independent of their total length and geometry.

---

[*] To whom correspondence should addressed. E-mail: engheta@ee.upenn.edu

As reported and discussed in [4]-[7], this energy squeezing is associated with a huge increase in the magnitude of the electric field inside the ENZ channels. Intuitively, one can expect that such high-intensity field may ensure high sensitivity to small variations in material parameters, which may then be exploited for material sensing applications. Here we analyze in details the perturbative effects of a small defect (e.g., a small change in material permittivity) or of a small cavity carved inside the metamaterial channel, with particular attention to the resulting changes in the channel tunneling properties. In particular, in the case of high-quality-factor (i.e., high-Q) tunneling, it is shown how the high-intensity electric field in the channel may be successfully used for accurate sensing applications and for fine tuning of the tunneling mechanism.

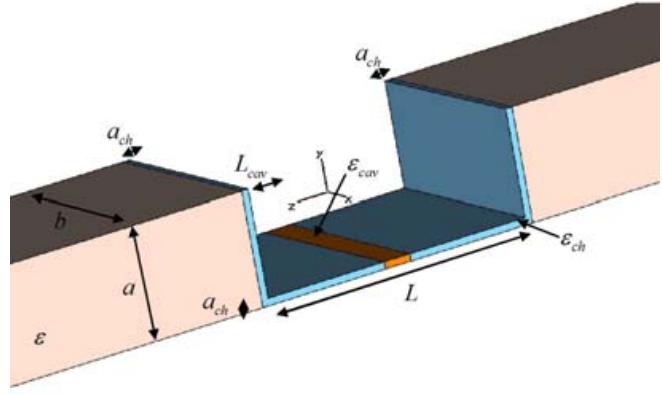

Figure 1 – (Color online) Geometry of the structure proposed as a permittivity sensor. The two input and output rectangular waveguides are connected via an ultranarrow channel (shown in blue). In the middle, an object (i.e., a cavity region) with permittivity $\varepsilon_{cav}$ different from that of the channel in inserted. The structure is all surrounded by conducting walls.

Consider the geometry of Fig. 1, consisting of an ultranarrow rectangular channel, of height $a_{ch}$ and length $L$, connecting the two sections of a waveguide of height $a \gg a_{ch}$. All the waveguide sections have the same width $b > a$, which tailors the dispersion properties of the fundamental $TE_{10}$ mode. In particular, the $TE_{10}$ propagation constant in each section of the waveguide may be regarded as that of a *TEM* wave propagating in an unbounded medium or in a parallel-plate waveguide with effective constitutive parameters [8]:

$$\varepsilon_{eff} / \varepsilon_0 = \varepsilon - c^2 / (4 f^2 b^2),$$
$$\mu_{eff} = \mu_0$$
(1)

where $\varepsilon$ is the permittivity of the homogeneous material filling the waveguide section, $c$ is the velocity of light in free-space and an $e^{-i2\pi ft}$ time convention has been assumed. At the cutoff frequency of this mode, i.e., when $b = \lambda/2$ with $\lambda$ being the wavelength in the filling material with permittivity $\varepsilon$, the effective permittivity (1) "experienced" by the mode is zero, consistent with the infinite phase velocity of a cutoff mode or of a *TEM* wave in an ENZ metamaterial. This direct analogy may be exploited to realize metamaterial effects in several geometries (see, e.g., [8]-[9]). Metamaterials realized with this technique are inherently robust to losses, since they avoid the use of small inclusions typical of conventional metamaterial technology, which usually lead to absorption and radiation losses associated with technological imperfections and disorder.

As we have done in [6], this technique may be applied to experimentally verify the anomalous properties of effective ENZ channels for resonant tunneling and zero-phase enhanced transmission, in principle independent of their geometry and length. In the geometry of Fig. 1 we have also considered two narrow sections of length $a_{ch}$ attached to the channel, with the aim of "matching" the incident mode into the narrow channel, forming a characteristic U shape for the ENZ region (blue region in Fig. 1).

In this way, we may obtain the ENZ-related supercoupling effects [4]-[6] using conventional materials, e.g., a dielectric or even free space with $\varepsilon_{ch} = \varepsilon_0$, filling the U channel. Since the ENZ-based tunneling is expected to occur at the cutoff frequency of the narrow channel, it is sufficient to have the remaining part of the waveguide filled with a regular dielectric with $\varepsilon > \varepsilon_{ch}$ to ensure propagation in these outer sections of the waveguide at this frequency. The phase delay between the entrance and the exit of the narrow channel is expected to remain very small, independent of the channel length, due to the infinite phase velocity of the mode near its cut-off, resulting in a uniform and strongly enhanced electric field along the channel.

Now we consider a small cavity region with length $L_{cav}$ and permittivity $\varepsilon_{cav} \neq \varepsilon_{ch}$ in the channel (Fig. 1). The presence of this cavity region with a permittivity different from the rest of the channel clearly perturbs the tunneling effect. In [6], we have studied the effects of an air cavity in the ENZ narrow channel on the electric and magnetic field distributions in the channel. Here we concentrate on how and to what extent the cavity with $\varepsilon_{cav} \neq \varepsilon_{ch}$ may alter and tune the tunneling and supercoupling aspects of the ENZ channel. As it will be shown in the following, this analysis may provide us with quantitative measures of the perturbation caused by the presence of a small object (here a cavity) of permittivity $\varepsilon_{cav}$. This may directly lead us to useful applications of this setup for sensing dielectric properties of an object under test.

The overall system of Fig. 1 may be analyzed using a transmission-line (TL) model, consistent with the one presented in [7] for a different geometry of the supercoupler, since each waveguide section supports the same dominant $TE_{10}$ mode [10]. In particular, each section is described by its characteristic impedance $\eta = \sqrt{\mu_0 / \varepsilon_{eff}}$ and wave number $\beta = 2\pi f \sqrt{\mu_0 \varepsilon_{eff}}$, and the steps in the U channel cross-section may be modeled with voltage transformers with a transform ratio $s = a_{ch} / a$ and parallel reactive loads that take into account the localized higher-order evanescent mode excitation. In this specific geometry, since the steps are embedded in an effectively ENZ region (the U channel) at the resonant frequency, these parallel loads may be neglected, since the condition $\varepsilon_{eff} \simeq 0$ implies an extremely high impedance for these loads [11]. Similarly, due to the condition $a_{ch} \ll 2\pi/\beta_{ch}$, it is possible to neglect the presence of the small lateral transition regions of the U in the TL model. We note that this same model (and the following discussions) applies for a parallel plate waveguide when filled with metamaterials with permittivity $\varepsilon_{eff}$, consistent with the geometry in [4].

In the absence of the cavity, or when $\varepsilon_{cav} = \varepsilon_{ch}$, the input impedance at the entrance of the channel may be written as:

$$Z_{in} = \frac{2s^2\eta_{ch}^2\eta_{out} - is\eta_{ch}(s\eta_{ch} - \eta_{out})(s\eta_{ch} + \eta_{out})\sin(2\beta_{ch}L)}{s^2\eta_{ch}^2 + \eta_{out}^2 + (s\eta_{ch} - \eta_{out})(s\eta_{ch} + \eta_{out})\cos(2\beta_{ch}L)}, \quad (2)$$

where the subscripts $_{ch}$ and $_{out}$ refer to the channel and the outside region, respectively. Total transmission is obtained when $Z_{in} = \eta_{out}$, i.e., either when $L = 2\pi/\beta_{ch}$ or when:

$$\eta_{ch} = \pm\frac{\eta_{out}}{s}. \quad (3)$$

The first solution corresponds to classic Fabry-Perot resonances of the channel, very sensitive to the length of the channel and the frequency of operation, whereas condition (3) is a unique matching condition obtainable only near the channel cutoff frequency, which is interestingly independent of its geometry (and in particular its length). The strong mismatch at the entrance and exit faces of the channel, due to the difference in heights, is compensated by the huge increase in its characteristic impedance for frequencies right above the cutoff. In the limit of $a_{ch} \to 0$, i.e., $s \to 0$, condition (3) requires the system to operate at the frequency for which $\varepsilon_{eff} = 0$, i.e., at the channel's cutoff frequency, independent of the channel length. This is consistent with the theoretical results obtained in this limiting case in [4] using a different approach. For finite, but narrow enough channels ($s \simeq 0$), Eq. (3) suggests operation at frequencies for which $\varepsilon_{eff} \simeq 0^+$. It should be mentioned that the steep variation of $\eta_{ch}$ right above the cutoff frequency implies that the tunneling frequency indeed occurs very close to the cutoff frequency, irrespective of variations in $a_{ch}$ (while $a_{ch} \ll a$), consistent with the theoretical and experimental results in [4]-[7]. The presence of the parallel loads, representing the E-plane discontinuities, would slightly shift down the tunneling frequency [11], but they do not notably affect the main features in this scenario. The low value of $\varepsilon_{eff}$ in the channel implies $\beta_{ch} \simeq 0$, which ensures small phase delay and uniform field inside the channel, whereas the continuity of voltage across the step leads to an enhancement of the electric field inside the channel by a factor of $1/s$ with respect to the field in the outer waveguide sections.

Consider now a small cavity region with permittivity $\varepsilon_{cav} \neq \varepsilon_{ch}$ in the narrow channel, as depicted in Fig. 1. Applying the same TL model, condition (3) is modified into:

$$\eta_{ch}^2\left(1+\frac{2\sin[\beta_{cav}L_{cav}](\eta_{cav}^2-\eta_{ch}^2)}{\begin{Bmatrix}\left[(\eta_{ch}^2-\eta_{cav}^2+(\eta_{ch}^2+\eta_{cav}^2)\cos[\beta_{ch}(L-L_{cav})]\right)\sin[\beta_{cav}L_{cav}]+\\ +2\eta_{cav}\eta_{ch}\cos[\beta_{cav}L_{cav}]\sin[\beta_{ch}(L-L_{cav})]\end{Bmatrix}}\right)=\frac{\eta_{out}^2}{s^2},$$

(4)

where the subscript $_{cav}$ refers to the cavity parameters. For the sake of simplicity, but without loss of generality, we have assumed that the cavity is positioned in the center of the channel. Eq. (4) quantifies the perturbation and the shift in the tunneling frequency, caused by the cavity. In particular, it may be deduced from (4) that the shift is always towards lower frequencies if $\varepsilon_{cav}>\varepsilon_{ch}$, implying that when the cavity is inserted the tunneling occurs at a frequency lower than the channel's cutoff frequency. It is interesting to point out that for this symmetric scenario, for which the cavity is placed at the center of the channel, total transmission can still be achieved when Eq. (4) is satisfied. For the asymmetric case, however, the resonant tunneling does not ensure, in general, total transmission.

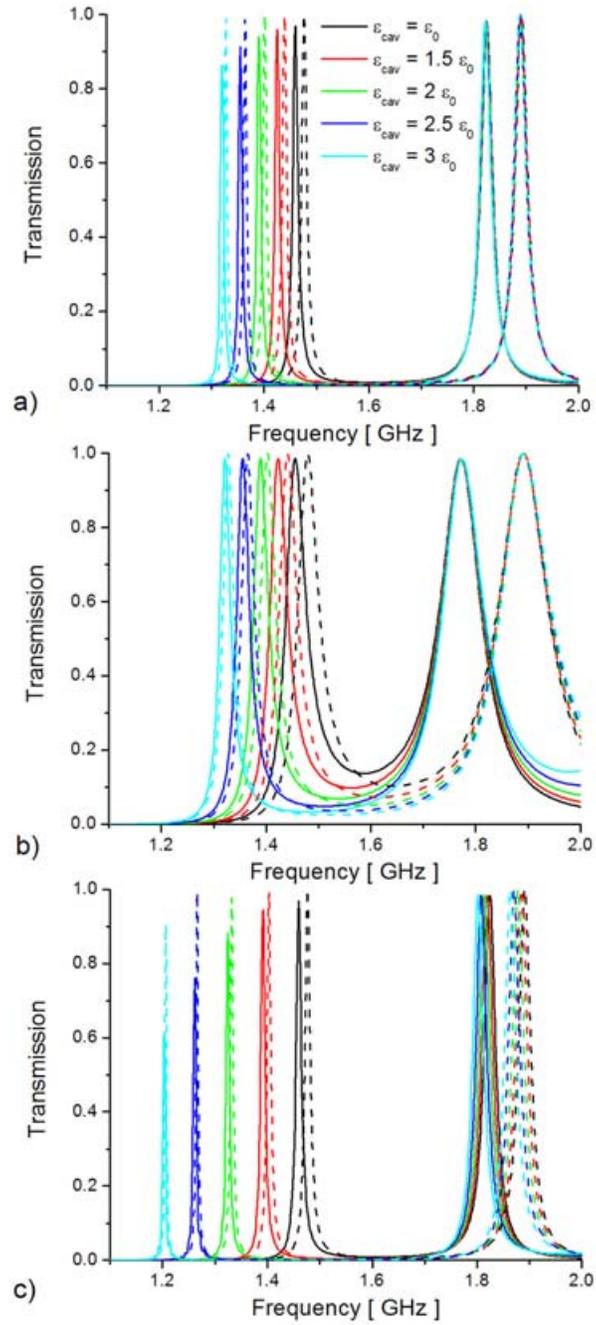

Figure 2 – (Color online) Transmission coefficient for the U channel of Fig. 1 with $L = 127\,mm$, $b = 2a = 101.6\,mm$, $\varepsilon = 2\varepsilon_0$, $\varepsilon_{ch} = \varepsilon_0$ and: (a) $a_{ch} = a/64$, $L_{cav} = L/10$; (b) $a_{ch} = a/16$, $L_{cav} = L/10$; (c) $a_{ch} = a/64$, $L_{cav} = L/5$. Solid lines: Full-wave simulations [12]; Dashed lines: TL model.

Figure 2 shows the transmission coefficient predicted using the analytical TL model described above (dashed line) and compared with the full-wave numerical simulations [12] (solid lines) for different geometries of the channel and the cavity. In all these examples $L = 127\,mm$, $b = 2a = 101.6\,mm$, $\varepsilon = 2\varepsilon_0$ (the permittivity of Teflon) and $\varepsilon_{ch} = \varepsilon_0$. The black lines refer to the case in which the cavity has the same permittivity as the channel, i.e., $\varepsilon_{cav} = \varepsilon_{ch} = \varepsilon_0$, which is consistent with the geometry analyzed in [6]. The ENZ-related tunneling occurs very close to the cutoff frequency of the channel, which happens at $f = 1.48\,GHz$ in this scenario. A second peak, consistent with a Fabry-Perot resonance is obtained at a higher frequency, but this is strongly dependent on the length $L$. Figure 2a considers the presence of a cavity with $L_{cav} = L/10$ in a channel with $a_{ch} = a/64$. Small variations in $\varepsilon_{cav}$ introduce a significant shift in the ENZ-related tunneling frequency, in accordance with Eq. (4). The full-wave numerical simulations agree very well with the analytical model derived above (which neglects the presence of reactive loads describing the steps, around this tunneling frequency), validating the assumption that the presence of an effectively ENZ region around the channel E-plane discontinuities significantly reduces the excitation of higher-order evanescent modes [11]. At a higher frequency, however, when the Fabry-Perot resonance arises (for this geometry around $1.8\,GHz$) and the channel has a larger positive $\varepsilon_{eff}$, the presence of these loads cannot be safely neglected, as revealed in the disagreement between the analytical and simulation lines around this frequency. We note here that the high-Q ENZ-related transmission peak in the channel may be fine tuned by varying the cavity permittivity $\varepsilon_{cav}$ around the tunneling frequency. However, the Fabry-Perot resonance peak, in comparison, is weakly affected by the presence of the cavity, due to the non-uniform distribution of electric field in the channel in the Fabry-Perot scenario. In particular, in this case the electric field distribution is such that the field is expected to be near zero in the center of the channel where the cavity is positioned.

Fig. 2b considers a similar geometry, but with a channel with higher height, i.e., $a_{ch} = a/16$. In this case the electric field enhancement in the channel is less, and correspondingly the Q of the tunneling transmission is also lowered, causing broadening of bandwidth of transmission peaks. Fig. 2c presents the results for a larger cavity, i.e., $L_{ch} = L/5$. In this case the resonance shift is more pronounced, as expected form the larger geometrical perturbation, but maintaining a similar Q factor as in the case of Fig. 2a, since $a_{ch}$ remains unchanged.

These results are summarized in Fig. 3, which reports the calculated shift of tunneling frequency (Fig. 3a) and the variation of the transmission coefficient at the channel cutoff frequency (Fig. 3b) for the three geometries of Fig. 2. This clearly reveals the filtering response of this ENZ-related tunneling phenomenon and suggests that the ENZ-inspired supercoupling may be exploited as a sensor for permittivity measurement by properly tailoring the tunneling Q factor. Operating at the channel's cutoff frequency, the transmission coefficient can be sensitively altered when a small object (particle, fluids, etc) with permittivity $\varepsilon_{cav}$ is inserted in the channel. By measuring this transmission coefficient, one can then evaluate the value of $\varepsilon_{cav}$, based on a curve such as Fig, 3b. We have assumed here that $\varepsilon_{cav}$ is a real quantity, since losses in dielectric materials may be safely neglected in this context, but this analysis remains valid also for absorbing materials.

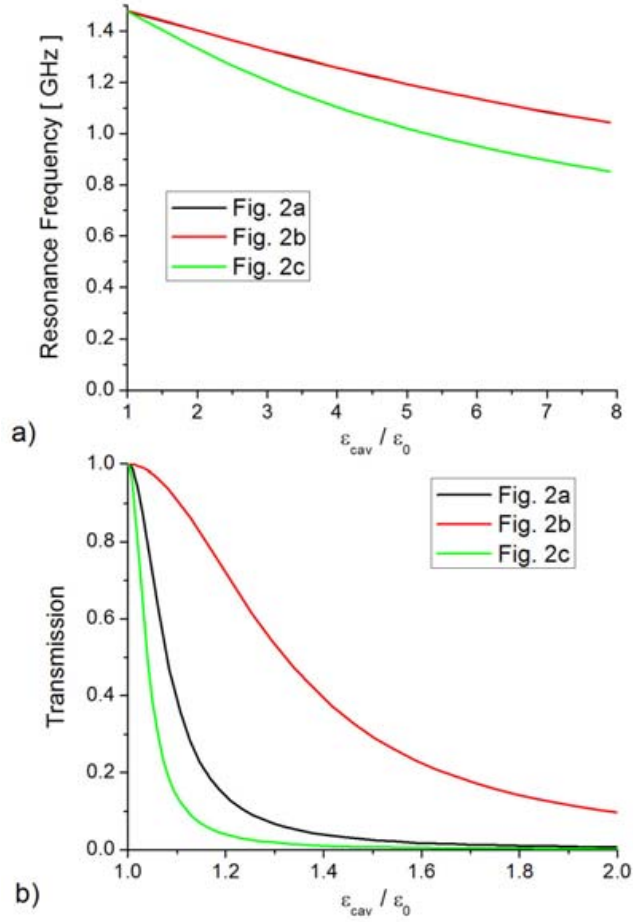

Figure 3 – (Color online) Variation of (a) the ENZ-related tunneling frequency and (b) the ENZ-related transmission coefficient at the channel's cutoff frequency $f = 1.48\,GHz$, vs. normalized $\varepsilon_{cav}$, for the geometries of Fig. 2.

Figure 4 shows the electric and magnetic field distributions, normalized to the impinging field amplitude, calculated on the bottom plate of the waveguide across the channel for the geometries of Fig. 2a and Fig. 2c with $\varepsilon_{cav} = 3\varepsilon_0$ at the corresponding resonant frequencies, as indicated in the caption. It is clear that we have a large increase of the normalized electric field distribution inside the cavity, even more than what is attainable within an empty supercoupling channel at its cutoff frequency, which for these geometries would correspond to a uniform distribution all across the channel with normalized amplitude equal to $s = 64$. With the cavity included, however, the electric field is further increased in the central region of the channel, where the cavity is placed, due to an anomalous growing exponential tail in the first part of the channel associated with resonant tunneling and the fact that the channel is resonating below its cutoff. This is even more evident in the magnetic field distribution (Fig. 4b), which would be constant in an empty ENZ channel [4]. The downshift of the tunneling frequency produces a large buildup of the normalized electromagnetic fields in the channel, localized at the cavity edges. In some sense, this is analogous to the exponential growth of the electromagnetic fields in the resonant pairs of metamaterials below cutoff [13], even if in this scenario the resonance interestingly happens in a simple hollow waveguide section. The field increase is more pronounced for a larger cavity, since the resonance happens at lower frequencies for which the effective permittivity of the channel is more negative, following Eq. (1).

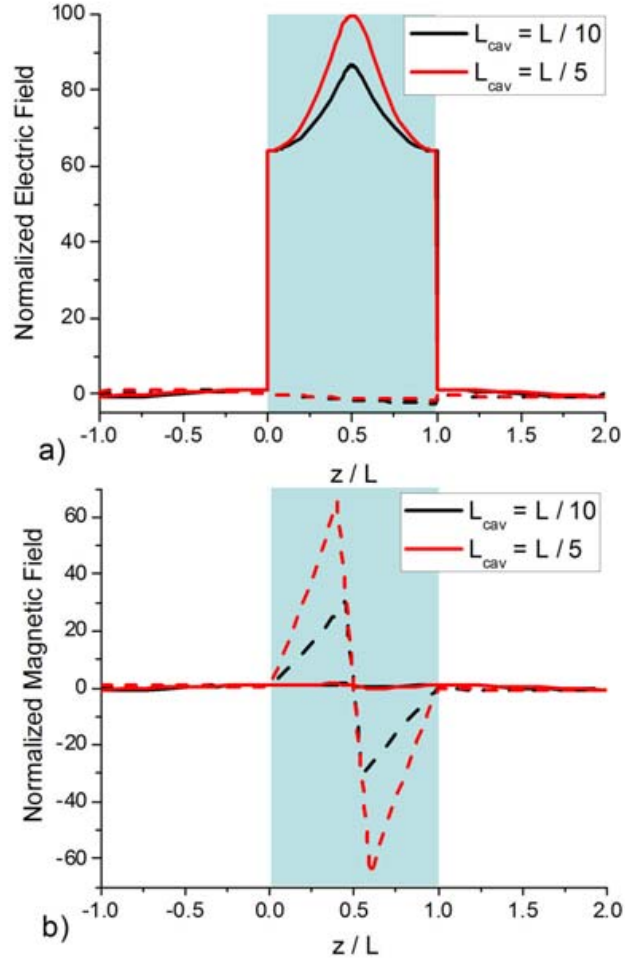

Figure 4 – (Color online) Normalized (a) electric $E_y$ and (b) magnetic $H_x$ field distributions [real (solid) and imaginary (dashed) part] on the bottom plate of the waveguide at the tunneling frequencies $f = 1.327\,GHz$ and $f = 1.206\,GHz$, respectively, for the geometries of Fig. 2a and Fig. 2c with $\varepsilon_{cav} = 3\varepsilon_0$. The channel region is highlighted in the plots.

The experimental microwave realization of dielectric sensing presented in this letter is being planned by our group, and the extension of these concepts to optical frequencies is also conceptually viable using plasmonic waveguide channels. This work is supported in part by the U.S. Office of Naval Research (ONR) grant number N 00014 -07-1-0622.